\documentclass[aps,showpacs,preprintnumbers,nofootinbibt,twocolumn, amsfonts]{revtex4}
\usepackage{graphicx}
\usepackage{amssymb}

\newcommand{\ba}{\begin{eqnarray}}
\newcommand{\ea}{\end{eqnarray}}
\newcommand{\be}{\begin{equation}}
\newcommand{\ee}{\end{equation}}
\newcommand{\lp}{\left(}
\newcommand{\rp}{\right)}

\begin{document}

\title{Palatini formulation of modified gravity with a nonminimal
curvature-matter coupling}

\author{Tiberiu Harko}
\email{harko@hkucc.hku.hk}
\affiliation{Department of Physics and
Center for Theoretical and Computational Physics, The University
of Hong Kong, Pok Fu Lam Road, Hong Kong, P. R. China}

\author{Tomi S. Koivisto}
\email{t.s.koivisto@uu.nl}
\affiliation{Institute for Theoretical Physics and Spinoza Institute, Utrecht University, 3508 Utrecht, Netherlands}

\author{Francisco S. N. Lobo}
\email{flobo@cii.fc.ul.pt} \affiliation{Centro de Astronomia e
Astrof\'{\i}sica da Universidade de Lisboa, Campo Grande, Ed. C8
1749-016 Lisboa, Portugal}

\begin{abstract}

We derive the field equations and the equations of motion for scalar fields and massive test particles in modified theories of gravity with an arbitrary coupling between geometry and matter by using the Palatini formalism.  We show that the independent connection can be expressed as the Levi-Civita connection of an auxiliary, matter Lagrangian dependent metric, which is related with the physical metric by means of a conformal transformation. Similarly to the metric case, the field equations impose the non-conservation of the energy-momentum tensor. We derive the explicit form of the equations of motion for massive test particles in the case of a perfect fluid, and the expression of the extra-force is obtained in terms of the matter-geometry coupling functions and of their derivatives. Generally, the motion is non-geodesic, and the extra force is orthogonal to the four-velocity. It is pointed out here that the force is of a different nature than in the metric formalism. We also consider the implications of a nonlinear dependence of the action  upon the matter lagrangian.

\end{abstract}


\date{\today}

\maketitle

\section{Introduction}

A promising way to explain the recent observational data \cite{Ri98,PeRa03} of the accelerated expansion of the Universe and of dark matter is to assume
that at large scales Einstein's general theory of relativity breaks
down, and a more general action describes the gravitational field \cite{review}. Theoretical models in which the standard Einstein-Hilbert action is replaced by an arbitrary function of the Ricci scalar $R$, first proposed in \cite{Bu70}, have been extensively investigated lately. For a review of $f(R)$ generalized gravity models and on their physical implications see \cite{SoFa08}. The possibility that the galactic dynamic of massive test particles can be understood without the need for dark matter was also considered in the framework of $\ f(R)$ gravity models \cite{darkmatter}.

A generalization of the $f(R)$ gravity theories was proposed in \cite
{Bertolami:2007gv} by including in the theory an explicit coupling of an
arbitrary function of the Ricci scalar $R$ with the matter Lagrangian
density $L_{mat}$ \cite{tomi}. As a result of the coupling the motion of the massive
particles is non-geodesic, and an extra force, orthogonal to the four-velocity, arises. The connections with Modified Orbital Newtonian Dynamics (MOND) and the Pioneer anomaly were also explored, and it was suggested that the matter-geometry coupling may be responsible for the observed behavior of the galactic rotation curves. In fact, the nonminimal curvature-matter coupling has received much attention lately. For instance, the model was extended to the case of the arbitrary couplings in both geometry and matter in \cite{ha08}. The implications of the non-minimal coupling on the stellar equilibrium were also investigated in \cite{Bertolami:2007vu}, where constraints on the coupling were obtained. An inequality which expresses a necessary and sufficient condition to avoid the Dolgov-Kawasaki instability for the model was derived in \cite{Fa07}. The relation between the model with geometry-matter coupling and ordinary scalar-tensor gravity, or scalar-tensor theories which include non-standard couplings between the scalar and matter was studied in \cite{SoFa08a}. In the specific case where both the action and the coupling are linear in $R$ the action leads to a theory of gravity which includes higher order derivatives of the matter fields without introducing more dynamics in the gravity sector \cite{So08}. The equivalence between a scalar theory and the model with the non-minimal coupling of the scalar curvature and matter was considered in \cite{BePa08}. This equivalence allows for the calculation of the Parameterized Post-Newtonian (PPN) parameters $\beta $ and $\gamma $, which may lead to a better understanding of the weak-field limit of $f(R)$ theories.

The equations of motion of test bodies in the nonminimal coupling model by means of a multipole method were also derived in \cite{Pu08}. Furthermore, the energy conditions and the stability of the model under the Dolgov-Kawasaki criterion were studied in \cite{Be09}. The perturbation equation of matter on subhorizon scales in models with an arbitrary matter-geometry coupling was used to constrain the theory from growth factor and weak lensing observations \cite{Ne09}. In particular, the age of the oldest star clusters and the primordial nucleosynthesis bounds were used in order to constrain the parameters of a toy model. The possibility that the behavior of the rotational velocities of test particles gravitating around galaxies can be explained in the framework of modified gravity models with non-minimal matter-geometry coupling was considered in \cite{Ha110}. Analogous nonlinear gravitational couplings with a matter Lagrangian were also considered in the context of proposals to address the cosmic accelerated expansion \cite{Odintsov}.

The subtle issue of a correct definition for the matter Lagrangian of the theory and of the definition of the energy-momentum tensor in the presence of a nonminimal curvature-matter coupling needs to be pointed out. In a recent paper \cite{Sotiriou:2008it}, the authors argued that a ``natural choice'' for the matter Lagrangian density for perfect fluids is $L_{mat}=p$, based on Refs. \cite{Schutz:1970my,Brown:1992kc}, where $p$ is the pressure. This choice has a particularly interesting application in the analysis of the curvature-matter coupling for perfect fluids, which imposes the vanishing of the extra force \cite{Bertolami:2007gv,Ha10}. Despite the fact that $L_{mat}=p$ does indeed reproduce the perfect fluid equation of state, it is not unique \cite{BLP}. Another choice includes, for instance, ${\cal L}_m=-\rho$ \cite{Brown:1992kc,HawkingEllis,Faraoni:2009rk}, where $\rho$ is the energy density (see Ref. \cite{BLP,Brown:1992kc} for details). For a review of modified $f(R)$ gravity with geometry-matter coupling see \cite{rev}.

In the literature different approaches in $f(R)$ modified theories of gravity are used. These include the metric formalism, where the action is varied with respect to the metric; the Palatini formalism, where the metric and the connections are treated as separate variables, and the Lagrangian is varied with respect to both to derive the field equations; and the metric-affine formalism, which generalizes the Palatini variation, where the matter part of the action depends and is varied with respect to the connection.
In this work, we use the Palatini formalism, that treats the metric and the affine connection as independent geometrical quantities. When the Einstein-Hilbert action is used, the Palatini variational principle leads to the Einstein equations, as in the standard metric variation. This is not true, however, for a more general action. When used together with an $f(R)$ Lagrangian, the Palatini formalism leads to second order differential equations instead of the fourth order ones that one gets with the metric variation. At the same time, in vacuum, they straightforwardly reduce to standard General Relativity (GR) plus a cosmological constant. This ensures us that, firstly, the theory passes the solar system tests, and secondly, that interesting aspects of GR like static black holes and gravitational waves are still present. Thus there is no real criterion so far about which one of them is better to use. Additionally, the Palatini variation seems to be more general since it yields GR without the need to specify the relation between the metric and the connection. The Palatini formalism for $f(R)$ modified gravity models has been intensively investigated recently \cite{all}.

As alternatives to dark energy, these models run into problems with structure formation \cite{structure}. While the modified gravity sector can support an accelerating universe without dark energy, the cosmological perturbations will then behave differently than in GR with a cosmological constant. In order to comply with the observations of the large-scale structure, one is then forced to to fine-tune the gravity lagrangian to be essentially GR with a cosmological constant. However, it has been shown that this can be avoided in the presence of generalized dark matter \cite{Koivisto:2007sq}. In particular, the evolution of cosmological perturbations can be viable if dark matter had effective pressures related to the modified gravity sector. Such effect ensues from a nonminimal coupling to curvature, motivating to study such a coupling particularly in the Palatini formalism. In fact, dark matter itself could emerge from a nonminimal coupling of baryons to curvature at supergalactic scales.

It is the purpose of the present paper to derive the gravitational field equations of the generalized $f(R)$ type gravity models with non-minimal coupling between matter and geometry, which depends on two arbitrary functions of the Ricci scalar and of the matter Lagrangian, respectively. By taking
two independent variations with respect to the metric and the connection
separately of the gravitational action, we obtain the field equations and the connection associated to the Ricci tensor, which, due to the matter-geometry coupling, is also a function of the matter Lagrangian. The metric that defines our independent connection is conformally related to the spacetime metric, with the conformal factor a function of the matter Lagrangian and of the Ricci scalar. Once the conformal factor is known, the field equations can be obtained easily in both metrics. By taking the divergence of the field equations it follows that the energy-momentum tensor of the matter is not conserved, and, similarly to the metric case,  due to the matter-geometry coupling an extra force arises.

The present paper is organized as follows. In Section \ref{2} the field equations of the modified gravity model with arbitrary matter-geometry coupling are obtained by using the Palatini formalism. The equation of motion of test bodies is derived in Section \ref{3}. We discuss and conclude our results in Section \ref{4}. In the present
paper we use the natural system of units with $8\pi G=c=1$.

\section{Field equations in Palatini modified gravity with an arbitrary coupling between matter and geometry}\label{2}

By assuming an arbitrary coupling between matter and geometry, the action for the modified theory of gravity takes the form \cite{Bertolami:2007gv, ha08}
\begin{eqnarray}\label{act}
S&=&\int \left\{ \frac{1}{2}f_{1}\left[R(g,\tilde{\Gamma })\right]+f_{2}\left[R(g,\tilde{\Gamma } )\right]G\left[
L_{mat}\left(g,\psi\right)\right]\right\}\times \nonumber\\
&&\times \sqrt{-g}\;d^{4}x~,
\end{eqnarray}
where $f_{i}(R)$ (with $i=1,2$) are arbitrary functions of the Ricci scalar $R=g^{\mu \nu}\tilde{R}_{\mu \nu}$, while $G\left(L_{mat}\right)$ is an arbitrary function of the matter Lagrangian density $L_{mat}$. The matter Lagrangian is a function of the metric tensor $g$ and of the physical fields $\psi$, while the Ricci tensor $\tilde{R}_{\mu \nu}$ is expressed solely in terms of the connection $\tilde{\Gamma }$. The only requirement for the
functions $f_i(R)$, $i = 1, 2$, and $G$, is to be analytical functions of the Ricci scalar $R$ and of the matter Lagrangian density $L_{mat}$, respectively, that is, they must possess a Taylor series expansion about any point \cite{ha08,pun}. The Ricci tensor is defined as \cite{LaLi, tomi}
\begin{equation}
\tilde{R}_{\mu \nu }=\partial _{\lambda }\tilde{\Gamma }_{\mu \nu }^{\lambda }-\partial
_{\nu }\tilde{\Gamma }_{\mu \lambda }^{\lambda }+\tilde{\Gamma }_{\mu \nu }^{\lambda }\tilde{\Gamma }
_{\lambda \alpha }^{\alpha }-\tilde{\Gamma }_{\mu \lambda }^{\alpha }\tilde{\Gamma }_{\nu
\alpha }^{\lambda },
\end{equation}%
with the connection $\tilde{\Gamma }_{\mu \nu }^{\lambda }$ obtained through the independent variation of the
gravitational field action given by Eq.~(\ref{act}), and not directly constructed from the metric by using the Levi-Civita prescription. The energy-momentum tensor of the matter is introduced as
\begin{equation}
T_{\mu \nu }=-\frac{2}{\sqrt{-g}}\frac{\delta \left( \sqrt{-g}L_{mat}\right)
}{\delta g^{\mu \nu }},
\end{equation}
thus giving
\begin{equation}
\frac{\delta L_{mat}}{\delta g^{\mu \nu }}=-\frac{1}{2}T_{\mu \nu }+\frac{1}{%
2}L_{mat}g_{\mu \nu }.
\end{equation}

The Palatini formalism consists in taking separately two independent variations with respect to the metric and the connection, respectively. The action is formally the same, but the Riemann tensor and the Ricci tensor are constructed with the independent connection. By varying the action (\ref{act}) with respect to the metric $g$ we obtain
\begin{eqnarray}\label{field1}
&&\left[ f_{1}^{\prime}(R)+2f_{2}^{\prime}(R)G\left( L_{mat}\right)
\right] \tilde{R}_{\mu \nu }-\nonumber\\
&&-\left\{ \frac{1}{2}f_{1}(R)+f_{2}(R)G\left(
L_{mat}\right) \left[ 1-L_{mat}\frac{G^{\prime }\left( L_{mat}\right) }{G\left(
L_{mat}\right) }\right] \right\}\times \nonumber\\
&&\times g_{\mu \nu }=f_{2}(R)G^{\prime }\left(
L_{mat}\right) T_{\mu \nu }\,,
\end{eqnarray}
where the prime denotes a derivative with respect to the argument, i.e., $f'_i(R)=df_i(R)/dR$ and $G^{\prime}\left( L_{mat}\right) =dG\left( L_{mat}\right)
/dL_{mat}$. In the limit $f_2\left(R\right)=1$, $G\left(L_{mat}\right)=L_{mat}$ we obtain the Palatini formalism field equations of the standard $f(R)$ modified theory of gravity \cite{SoFa08,tomi}. In the case $f_1(R)=R$, $f_2(R)=1$, and $G\left(L_{mat}\right)=L_{mat}$, we obtain the field equations of standard general relativity.

The next step in the Palatini formalism requires the variation of the action with respect to the connection $\Gamma $. The variation can be done by using the identity
\begin{equation}
\delta \tilde{R}_{\mu \nu }=\tilde{\nabla} _{\lambda }\left( \delta \tilde{\Gamma }_{\mu \nu
}^{\lambda }\right) -\tilde{\nabla }_{\mu }\left( \delta \tilde{\Gamma }_{\nu \lambda}^{\lambda }\right) ,
\end{equation}%
where $\tilde{\nabla }_{\lambda }$ is the covariant derivative associated with the connection $\tilde{\Gamma}$.

By taking the variation of the action (\ref{act}) with respect to the
connection $\tilde{\Gamma }$ we obtain
\begin{equation}
\frac{\delta S}{\delta \tilde{\Gamma }}=\frac{1}{2}\int A^{\mu \nu }\left[ \tilde{\nabla }
_{\lambda }\left( \delta \tilde{\Gamma }_{\mu \nu }^{\lambda }\right) -\tilde{\nabla }_{\mu
}\left( \delta \tilde{\Gamma }_{\nu \lambda }^{\lambda }\right) \right] \sqrt{-g}%
d^{4}x,
\end{equation}%
where we have denoted
\begin{equation}
A^{\mu \nu }=\left[ \frac{1}{2}f_{1}^{\prime }(R)+f_{2}^{\prime }(R)G\left(
L_{mat}\right) \right] g^{\mu \nu }.
\end{equation}

By integrating by parts we obtain
\begin{eqnarray}
\frac{\delta S}{\delta \tilde{\Gamma }}&=&\frac{1}{2}\int \tilde{\nabla }_{\lambda }\left[
\sqrt{-g}\left( A^{\mu \nu }\delta \tilde{\Gamma }_{\mu \nu }^{\lambda }-A^{\lambda
\nu }\delta \tilde{\Gamma }_{\nu \alpha }^{\alpha }\right) \right] d^{4}x+\nonumber\\
&&+\frac{1}{2}%
\int \tilde{\nabla }_{\mu }\left[ \sqrt{-g}\left( A^{\mu \nu }\delta _{\alpha
}^{\lambda }-A^{\lambda \nu }\delta _{\alpha }^{\mu }\right) \right] \delta
\tilde{\Gamma }_{\lambda \nu }^{\alpha }d^{4}x.
\end{eqnarray}

The first term in $\delta S/\delta \tilde{\Gamma }$  is a total derivative, and thus it can be discarded. Hence the variation of the action with respect to the
connection gives
\begin{equation}
\tilde{\nabla }_{\mu }\left[ \sqrt{-g}\left( A^{\mu \nu }\delta _{\alpha }^{\lambda
}-A^{\lambda \nu }\delta _{\alpha }^{\mu }\right) \right] =0.  \label{con}
\end{equation}
Equation (\ref{con}) can be further simplified if we take into account that when $\alpha =\lambda $ the equation is identically zero. Taking $\lambda \neq \alpha $, we obtain
\begin{equation}\label{con1}
\tilde{\nabla }_{\alpha }\left\{ \sqrt{-g}\left[ f_{1}^{\prime }(R)+2f_{2}^{\prime
}(R)G\left( L_{mat}\right) \right] g^{\mu \nu }\right\} =0.
\end{equation}
Equation (\ref{con1}) shows that the connection is compatible with a conformal metric. By defining, according to \cite{SoFa08,tomi}, a new metric $h_{\mu \nu}$, conformal to $g_{\mu \nu }$, given by
\begin{equation}
h_{\mu \nu
}\equiv \left[ f_{1}^{\prime }(R)+2f_{2}^{\prime }(R)G\left( L_{mat}\right) %
\right] g_{\mu \nu },
\end{equation}
 we obtain
\begin{equation}
\sqrt{-h}h^{\mu \nu }=\sqrt{-g}\left[
f_{1}^{\prime }(R)+2f_{2}^{\prime }(R)G\left( L_{mat}\right) \right] g^{\mu
\nu },
\end{equation}
where $h$ is the determinant of the metric $h_{\mu \nu}$. Thus Eq.~(\ref{con1}) becomes the definition of the Levi-Civita
connection $\tilde{\Gamma }$ of $h_{\mu \nu }$, giving
\begin{equation}
\tilde{\Gamma }_{\mu \nu }^{\lambda }=\frac{1}{2}h^{\lambda \rho }\left(
\partial _{\nu }h_{\mu \rho }+\partial _{\mu }h_{\nu \rho }-\partial _{\rho
}h_{\mu \nu }\right) .
\end{equation}

With the use of the explicit form of $h_{\mu \nu }$ we obtain
\begin{equation}\label{con2}
\tilde{\Gamma }_{\mu \nu }^{\lambda }=\frac{1}{2}\frac{g^{\lambda \rho }}{F}%
\left[ \partial _{\nu }\left( Fg_{\mu \rho }\right) +\partial _{\mu }\left(
Fg_{\nu \rho }\right) -\partial _{\rho }\left( Fg_{\mu \nu }\right) \right],
\end{equation}
where we have denoted
\begin{equation}
F=F\left(R,L_{mat}\right)=f_{1}^{\prime }(R)+2f_{2}^{\prime }(R)G\left(
L_{mat}\right).
\end{equation}

In terms of the Levi-Civita connection $\Gamma _{\mu \nu}^{\lambda }$ associated to the metric $g$,
\begin{equation}\label{con3}
\Gamma _{\mu \nu }^{\lambda }=\frac{1}{2}g^{\lambda \rho }\left( \partial
_{\nu }g_{\mu \rho }+\partial _{\mu }g_{\nu \rho }-\partial _{\rho }g_{\mu
\nu }\right),
\end{equation}
$\tilde{\Gamma }_{\mu \nu }^{\lambda }$ can be expressed as
\begin{equation}
\tilde{\Gamma}_{\mu \nu }^{\lambda }=\Gamma _{\mu \nu }^{\lambda }+\partial
_{\nu }\ln \sqrt{F}\delta _{\mu }^{\lambda }+\partial _{\mu }\ln \sqrt{F}%
\delta _{\nu }^{\lambda }-g_{\mu \nu }g^{\lambda \rho }\partial _{\rho }\ln
\sqrt{F}.
\end{equation}

The Ricci tensor $\tilde{R}_{\mu \nu}$ is given in terms of the tensor $R_{\mu \nu}$ constructed from the metric by using the Levi-Civita connection Eq.~(\ref{con3}) by \cite{SoFa08}
\begin{eqnarray}\label{ricci}
\tilde{R}_{\mu \nu }&=&R_{\mu \nu }(g)+\frac{3}{2}\frac{1}{F^{2}}\left( \nabla
_{\mu }F\right) \left( \nabla _{\nu }F\right) \nonumber\\
&&-\frac{1}{F}\left( \nabla _{\mu }\nabla _{\nu }-\frac{1}{2}g_{\mu \nu } \Box \right)F ,
\end{eqnarray}%
while  the Ricci scalar and the Einstein tensor can be obtained as
\begin{equation}
\tilde{R}=R\left( g\right) +3\frac{1}{F}\square F+\frac{3}{2}\frac{1}{F^{2}}%
\left( \nabla _{\mu }F\right) \left( \nabla ^{\mu }F\right) ,
\end{equation}
and
\begin{eqnarray}\label{ein}
\tilde{G}_{\mu \nu }&=&\tilde{R}_{\mu \nu}-\frac{1}{2}g_{\mu \nu }\tilde{R}=G_{\mu \nu }(g)+\frac{3}{2}\frac{1}{F^{2}}\left( \nabla
_{\mu }F\right) \left( \nabla _{\nu }F\right) \nonumber\\
&&\hspace{-1.0cm}-\frac{1}{F}\left( \nabla
_{\mu }\nabla _{\nu }+g_{\mu \nu }\square \right)F -\frac{3}{4}\frac{1}{%
F^{2}}g_{\mu \nu }\left( \nabla _{\lambda }F\right) \left( \nabla ^{\lambda
}F\right) ,
\end{eqnarray}
respectively, with all the covariant derivatives taken with respect to the metric $g_{\mu \nu}$.

With the use of the expression of the Einstein tensor given by Eq.~(\ref{ein}) the gravitational field equation Eq.~(\ref{field1}) can be written as
\begin{eqnarray}\label{fieldp}
\tilde{G}_{\mu \nu }&=&\left( \frac{2K}{F}-\frac{1}{2}R-\frac{3}{2}\frac{1}{F}%
\square F-\frac{3}{4}\frac{1}{F^{2}}\nabla _{\lambda } F\nabla ^{\lambda
} F\right) g_{\mu \nu }\nonumber\\
&&+\frac{f_{2}\left( R\right) G\left( L_{mat}\right) }{F}%
T_{\mu \nu }\,,
\end{eqnarray}
where for notational simplicity, we introduced the notation
\begin{eqnarray}
K=K\left( R,L_{mat}\right) &=&
\left\{ f_{1}(R)+2f_{2}(R)G\left( L_{mat}\right)\right. \times\nonumber\\
&&\left.\times\left[ 1-L_{mat}\frac{G^{\prime }\left( L_{mat}\right) }{G\left(
L_{mat}\right) }\right] \right\} ,
\end{eqnarray}

By substituting the expression of the Einstein tensor given by Eq.~(\ref{ein}) into the field equation Eq.~(\ref{fieldp}) we obtain the gravitational field equation of modified gravity with a nonminimal coupling between matter and geometry in the Palatini formalism as
\begin{eqnarray}\label{fieldf}
&&G_{\mu \nu }+\frac{3}{2}\frac{1}{F^{2}}\left( \nabla _{\mu }F\right) \left(
\nabla _{\nu }F\right) -\frac{1}{F}\nabla _{\mu }\nabla _{\nu }F=\nonumber\\
&&\hspace{-0.75cm}\left( \frac{2K}{F}-\frac{1}{2}\frac{1}{F}\square F-\frac{R}{2}\right) g_{\mu \nu }+\frac{f_{2}\left( R\right) G\left( L_{mat}\right) }{F}T_{\mu \nu }.
\end{eqnarray}

Note that the field equation Eq.~(\ref{field1}) can be written as
\begin{equation}
F\tilde{R}_{\mu \nu }-\frac{1}{2}K\left( R,L_{mat}\right) g_{\mu \nu
}=f_{2}(R)G^{\prime }\left( L_{mat}\right) T_{\mu \nu }.  \label{field3}
\end{equation}
By taking the trace of Eq.~(\ref{field3}) we obtain
\begin{equation}\label{trace}
F\tilde{R}-2K\left( R,L_{mat}\right)\nonumber\\
=f_{2}(R)G^{\prime }\left( L_{mat}\right) T,
\end{equation}
where $T=T_{\mu }^{\mu }$ is the trace of the energy-momentum tensor. With
the use of Eq.~(\ref{ricci}) we find the equation determining $R$ as a function of $T$ and the matter Lagrangian $L_{mat}$ as
\begin{eqnarray}
&&R\left( g\right) -3\frac{1}{F}\square F-\frac{3}{2}\frac{1}{F^{2}}\left(
\nabla _{\mu }F\right) \left( \nabla ^{\mu }F\right) \nonumber\\
&&-2\frac{1}{F}K\left( R,L_{mat}\right) =
\frac{f_{2}(R)G^{\prime }\left( L_{mat}\right) }{F}T.
\end{eqnarray}

\section{Equation of motion and the extra force}\label{3}

\subsection{Non-geodesic motion}

By taking the covariant divergence of Eq.~(\ref{field3}) we obtain
\begin{eqnarray}
&&\left( \nabla ^{\mu }F\right) \tilde{R}_{\mu \nu }+F\nabla ^{\mu }\tilde{R}%
_{\mu \nu }-\frac{1}{2}g_{\mu \nu }\nabla ^{\mu }K\left( R,L_{mat}\right) =\nonumber\\
&&=\left[ \nabla ^{\mu }f_{2}(R)G^{\prime }\left( L_{mat}\right) \right] T_{\mu
\nu }+f_{2}(R)G^{\prime }\left( L_{mat}\right) \nabla ^{\mu }T_{\mu \nu }.\nonumber\\
\label{cons1}
\end{eqnarray}

With the use of the identity \cite{tomi}
\begin{equation}
\nabla ^{\mu }\tilde{G}_{\mu \nu }=-\frac{\nabla ^{\mu }F}{F}\tilde{R}_{\mu
\nu },
\end{equation}
we can write
\begin{equation}
F\nabla ^{\mu }\tilde{R}_{\mu \nu }+\left( \nabla ^{\mu }F\right) \tilde{R}%
_{\mu \nu }=\frac{1}{2}g_{\mu \nu }F\nabla ^{\mu }\tilde{R}.
\end{equation}
Using this in Eq.(\ref{cons1}) we then obtain that
\ba
\nabla_\mu T^\mu_{\phantom{\mu}\nu}& = & \frac{1}{f_2(R)G'(L_{mat})}\Big\{ - \lp\nabla_\mu f_2(R)G'(L_{mat})\rp T^{\mu}_{\phantom{\mu}\nu}
\nonumber \\ & + & L_{mat}G'(L_{mat})\nabla_\nu f_2(R)
 -  f_2(R)\nabla_\nu \mathcal{G}\Big\}\,,
\ea
where for notational convenience we introduced the shorthand $\mathcal{G} = G(L_{mat})-L_{mat}G'(L_{mat})$, which vanishes when the action
is linear in the matter lagrangian.



For a perfect fluid, described by an energy density $\epsilon$ and a thermodynamic pressure $p$,
the matter energy-momentum tensor is given by
\begin{equation}
T_{\mu \nu }^{(m)}=\left( \epsilon +p\right) u_{\mu }u_{\nu
}+pg_{\mu \nu },
\end{equation}
where the four-velocity, $u_{\mu }$, satisfies
the conditions $u_{\mu }u^{\mu }=-1$ and $u^{\mu }u_{\mu ;\nu }=0$. The trace of the energy-momentum tensor is given by $T=3p-\epsilon$.
We also introduce the projection operator $h_{\mu \lambda }=g_{\mu
\lambda }+u_{\mu }u_{\lambda }$ from which one obtains $h_{\mu
\lambda }u^{\mu }=0$ and $T^{\mu \nu } h_{\mu \lambda }= ph^{\nu}_{\lambda }$, respectively.

By contracting Eq. (\ref{cons1}) with the projection operator
$h_{\mu \lambda }$, one deduces the following expression
\begin{eqnarray}
&&\left( \epsilon +p\right) g_{\mu \lambda }u^{\nu }\nabla _{\nu }u^{\mu
}-(\nabla _{\nu }p)(\delta _{\lambda }^{\nu }-u^{\nu }u_{\lambda })= \nonumber\\
&&\frac{1}{f_{2}(R)G^{\prime }\left( L_{mat}\right) }\left( \delta _{\lambda }^{\nu }-u^{\nu }u_{\lambda
}\right)\times \Big\{ p\nabla_\nu \lp f_{2}(R)G^{\prime }\left( L_{mat}\right)\rp \nonumber\\
&& L_{mat}G'(L_{mat})\nabla_\nu f_2(R) -  f_2(R)\nabla_\nu \mathcal{G} \Big\}\,.
\end{eqnarray}
Finally, contraction with $g^{\alpha \lambda }$ gives rise to the equation
of motion for a fluid element
\begin{equation}
\frac{Du^{\alpha }}{ds}\equiv \frac{du^{\alpha }}{ds}+\Gamma _{\mu \nu
}^{\alpha }u^{\mu }u^{\nu }=f^{\alpha }~,
\end{equation}
where we have introduced the space-time connection $\Gamma _{\mu \nu
}^{\alpha }$, which is expressed in terms of the Christoffel symbols
constructed from the metric $g_{\mu \nu }$,  and where the extra-force is defined as
(dropping the explicit dependence of $G$ upon $L_{mat}$):
%
\ba
f^{\alpha }&=&\frac{h^{\nu \alpha }}{\epsilon +p}\Big[-\nabla_\nu p - \lp p-L_{mat}\rp \nabla_\nu  \log{f_2(R)} \nonumber \\
& + &  p \nabla_\nu\log{G'} + \frac{1}{G'}\nabla_\nu \lp G-L_{mat}G'\rp \Big]\,. \label{force}
\ea
As one can see from Eq. (\ref{force}), the extra force $f^{\alpha }$ is
orthogonal to the four-velocity of the particle,
\begin{equation}
f^{\alpha }u_{\alpha }=0~.
\end{equation}
The extra force-four velocity orthogonality relation follows from the properties
of the projection operator. This is consistent with the usual interpretation
of the force, according to which only the component of the four-force that
is orthogonal to the particle's four-velocity can influence its trajectory.
The presence of
the extra force $ f^{\alpha }$ implies that the motion of the
particle is non-geodesic. For $f^{\alpha }\equiv 0$ we recover the
geodesic equation of motion. The usual gravitational effects, due
to the presence of an arbitrary mass distribution, are assumed to
be contained in the term $a_{N}^{\mu }=\Gamma _{\alpha \beta
}^{\mu }u^{\alpha }u^{\beta }$.

We can now separate three contributions to the force. The first term in (\ref{force}) is the usual which is due to the gradient of pressure. The second term is due to the nonminimal curvature coupling to gravity and of course vanishes in general relativity. We note that the result reached in the metric formalism applies also here, that when $L_{mat}=p$, the extra force in fact disappears. Interestingly, this is the case in the usual lagrangian description of scalar fields and can hold for more general perfect fluids as well, depending upon the chosen prescription.

Furthermore, the two following terms, consisting of the last line in (\ref{force}), are due to the new self-interactions of matter and disappear when the action is linear in the matter lagrangian $L_{mat}$.  All of the three terms can be considered as contact forces. For a test particle, these vanish when in vacuum. This reflects the fact that the Palatini theory of gravity can be understood as a modified response to matter sources whereas it reduces to general relativity (with a cosmological constant) in vacuum. In other words, though the geodesics of particles are modified within a matter distribution, there are no new propagating degrees of freedom which would mediate interactions in vacuum and the extra force is absent there.

\subsection{Scalar field}

Let us as an example consider a general scalar field theory given by a lagrangian depending upon the field $\phi$.
The Euler-Lagrange equations of motion read then
\ba
&&\nabla_\mu\lp \frac{\partial L_{mat}}{\partial\lp \nabla_\mu\phi\rp}\rp - \frac{\partial L_{mat}}{\partial\phi} \nonumber \\ &=&
-\nabla_\mu\lp\log{f_2(R)G'(L_{mat})}\rp\frac{\partial L_{mat}}{\partial\lp \nabla_\mu\phi\rp}\,.
\ea
The left hand side represents now the extra force. This is due to nonminimal curvature coupling if $f_2(R)$ is not a constant, and due to matter self interactions, if $G'(L_{mat})$ is not a constant. Consider the purely latter case. As found above, now the extra-force entering the geodesic equation vanishes, but however there are additional terms in the Klein-Gordon equation for the field in curved space. Thus the additional terms do not affect the motion of a scalar particle. On the other hand, in the latter case, we obviously obtain new effects already in Minkowski space, and these can be regarded as nongravitational extra force.

As an explicit example, we let us consider an action given by an exponential of the Einstein-Hilbert lagrangian. This we fix the exponential dependence of the nonminimally coupled matter sector as
\ba
f_2(R) & = & \exp\lp{\frac{R}{2\kappa^2 \Lambda}}\rp\,,  \\
G(L_{mat}) & = & \Lambda\exp\lp{\frac{L_{mat}}{\Lambda}}\rp\,,
\ea
where $\Lambda$ is a suitable energy scale. Let us keep the form of the scalar field lagrangian as in the canonical theory,
\be
L_{mat} = \frac{1}{2}\lp \partial \phi\rp^2 - V(\phi)\,.
\ee
The Klein-Gordon equation then becomes
\ba
\Box\phi-V'(\phi)+ \frac{1}{\Lambda}\Big[\partial^\alpha\phi\partial^\beta\phi\nabla_\alpha\partial_\beta\phi+V'(\phi)\lp\partial\phi\rp^2 \nonumber \\
-\frac{1}{2\kappa^2}\partial_\mu R\partial^\mu\phi\Big] = 0 \,.
\ea
The corrections appear within the square brackets and are suppressed by the scale $\Lambda$. The propagator is modified in a nonlinear way. Note that only the last term in the second line drops when we consider flat instead of curved space.

\section{Discussions and final remarks}\label{4}

In the present paper we have considered the gravitational field equations for a generalized $f(R)$ type gravity model with a geometry-matter coupling in the framework of the Palatini formalism. Similarly to the metric case \cite{Bertolami:2007gv, ha08} the energy-momentum tensor of the matter is not conserved.  Palatini type $f(R)$ theories have special properties that make them especially interesting for addressing strong gravity phenomena
such as the early Universe or stellar collapse processes \cite{all}.
The coupling of the matter Lagrangian with the curvature term leads to an extra-force term in the equation of motion of massive particles in gravitational fields, which essentially depends on the allowed functional forms for the geometry-matter coupling. If, for example, the geometry-matter coupling could generate some repulsive forces, then one could construct non-singular cosmological models that are non-singular at high curvatures, or even non-singular collapsing stars.  No new degrees of freedom in the gravitational side or in the matter side (exotic sources) are needed to get such repulsive gravitational forces, since the extra-force is induced by the coupling between matter and geometry. Our results suggest that Palatini type theories might play a relevant role in the phenomenology of gravitation at both high energies (densities), as well as in the very low density limit. On the other hand, the assumption of independence between metric and connection in the variational process is essential to get second order equations for the
metric. It is thus reasonable to expect that effective descriptions of the gravitational force at large/small scales could come in the form of Palatini theories.

In three dimensions and in the Newtonian limit, Eq.~(\ref{force}) can be formally represented as a three-vector equation of the form $\vec{a}=\vec{a}_{N}+\vec{f}$ where $\vec{a}$ is the total acceleration of the particle, $\vec{a}_{N}$ is the gravitational acceleration and $\vec{f}$ is the acceleration (per unit mass) due to the presence of the extra force. Similarly to the metric case \cite{Bertolami:2007gv}, a Modified Newtonian Orbital Dynamics (MOND) \cite{mond} type acceleration is induced by the presence of the geometry-matter coupling. This shows that one possibility of testing the effects of the matter-geometry coupling could be in the region of very small accelerations of the order of $10^{-10}$ m/s$^2$. A supplementary MOND-type acceleration can explain the observed behavior of test particles gravitating around galaxies.

As a possible physical application of the Palatini formalism gravitational field equations one could consider an alternative view to the dark matter problem, namely, the possibility that the galactic rotation curves, and the mass discrepancy in galaxies and clusters of galaxies, which is usually explained by postulating the existence of dark matter,  can be explained in gravitational models in which there is a non-minimal coupling between
matter and geometry \cite{Bertolami:2007gv, ha08, Ha110}. Similarly to the metric case, the extra-terms in the gravitational field equations of the Palatini formalism modify the equations of motion of test particles, and induce a supplementary gravitational interaction, which can account for the observed behavior of the galactic rotation curves. Therefore, the analysis of the motion of the test particles at a galactic scale could open the possibility of directly testing the modified gravity models with non-minimal coupling between matter and geometry in the Palatini formalism by using direct astronomical and astrophysical observations at the galactic or extra-galactic scale. In this paper we have provided some basic theoretical tools necessary for the in depth comparison of the predictions of this model and of the observational results.

\section*{Acknowledgments}

The work of TH is supported by an RGC grant of the government of the Hong Kong SAR. TH would like to thank the
Centro de Astronomia e Astrof\'{\i}sica da Universidade de Lisboa for their support and warm hospitality during
the preparation of this work. TSK is supported by FOM and the Academy of Finland. FSNL acknowledges financial
support of the Funda\c{c}\~{a}o para a Ci\^{e}ncia e Tecnologia through the grants
PTDC/FIS/102742/2008 and CERN/FP/109381/2009.

\end{document}